\documentclass[letterpaper, conference]{IEEEtran}

\usepackage{cite}
\usepackage{amsmath,amssymb,amsfonts}
\usepackage{algorithmic}
\usepackage{graphicx}
\usepackage{textcomp}
\usepackage{xcolor}
\usepackage[utf8]{inputenc}
\usepackage{subcaption}
\usepackage{url}

\begin{document}

\title{Evaluating the Q-score of Quantum Annealers}
\author{Ward van der Schoot, Daan Leermakers, Robert Wezeman, Niels Neumann, Frank Phillipson}
\author{\IEEEauthorblockN{Ward van der Schoot}
\IEEEauthorblockA{TNO, The Netherlands} \and \IEEEauthorblockN{Daan Leermakers}
\IEEEauthorblockA{TNO, The Netherlands} \and \IEEEauthorblockN{Robert Wezeman}
\IEEEauthorblockA{TNO, The Netherlands} \and \IEEEauthorblockN{Niels Neumann}
\IEEEauthorblockA{TNO, The Netherlands} \and \IEEEauthorblockN{Frank Phillipson}
\IEEEauthorblockA{TNO, The Netherlands and Maastricht University, The Netherlands}}

\author{\IEEEauthorblockN{Ward van der Schoot\IEEEauthorrefmark{1}\IEEEauthorrefmark{3}, Daan Leermakers, Robert Wezeman\IEEEauthorrefmark{1}, Niels Neumann\IEEEauthorrefmark{1} and Frank Phillipson\IEEEauthorrefmark{1}\IEEEauthorrefmark{2}}\\
\IEEEauthorblockA{\IEEEauthorrefmark{1} The Netherlands Organisation for Applied Scientific Research,\\
Anna van Buerenplein 1, 2595DA, The Hague, The Netherlands}
\IEEEauthorblockA{\IEEEauthorrefmark{2} Maastricht University, School of Business and Economics,\\
P.O. Box 616, 6200 MD Maastricht, The Netherlands}
\IEEEauthorblockA{\IEEEauthorrefmark{3} Email: ward.vanderschoot@tno.nl}}

\date{March 2022}

\maketitle

\begin{abstract}
We report the Atos Q-score for D-Wave's quantum devices, classical algorithms and hybrid quantum-classical solver. Computing the Q-score entails solving the Max-Cut problem for increasingly large graphs. This work presents the first computation of the Q-score on a quantum device and shows how these quantum devices compare to classical devices at solving optimisation problems. We use D-Wave's standard methods out of the box with a time limit of 60 seconds.
The Q-score for D-Wave's 2000Q and Advantage devices are 70 and 140, respectively. The Q-score for two of D-Wave's classical algorithms, based on tabu search and simulated annealing respectively, are 2,300 and 5,800. Finally, we report the out-of-the-box hybrid approach to have a Q-score of 12,500.
\end{abstract}

\begin{IEEEkeywords}
Quantum computing, application-oriented benchmarks, quantum annealing, Q-score, D-Wave, Max-Cut.
\end{IEEEkeywords}

\section{Introduction}
    We live in interesting times. General purpose quantum computers, once a theoretical concept, are getting closer to outperforming classical computers on certain tasks. Many different parties are trying to build general-purpose quantum computers, such as IBM \cite{IBM} and Atos \cite{Atos}. At the same time quantum annealing devices promise utility in the shorter term in the domain of optimisation problems. Hybrid systems consisting of classical computing and quantum annealing are already being used to solve real world problems today \cite{hybridvanNiels}, \cite{feld2019hybrid}, \cite{phillipson2021portfolio}, \cite{TNO_Frank}.

Interest in these devices from both investors and end users is growing rapidly. To help these parties compare the performance of different quantum devices and accelerate progress, we require \textit{performance benchmarks}. Such metrics show the absolute and relative capabilities of a certain device with respect to some method chosen by the benchmark. A good benchmark is i) useful: it tells interested parties what they need to know, ii) hardware-agnostic: it does not favour a certain hardware implementation over another, iii) scalable: it can adapt to developments in quantum devices, and iv) comprehensive: it measures the quantum set-up as a whole, including the classical parts\footnote{We note here that benchmarks that do not have properties (iii) and (iv) can still be useful in the short term.} \cite{BCG}. In addition, it is useful for a benchmark to have an easy representation, to make it understandable and allow easy comparison of the performance of different devices. An example would be to have a benchmark given by a single number.

Originally, the number of qubits was the leading benchmark for quantum computing. However, it turns out that this metric does not tell the whole story, as it is neither useful, nor hardware-agnostic, nor comprehensive. This can for example be seen by comparing the number of qubits of D-Wave's latest quantum annealer (5640) \cite{Tech_report_dwave} and IBM's largest gate-based quantum computer (127) \cite{IBM-blog}. Since then, different parties have come up with new metrics, but these new metrics are not always adapted by other vendors. Different metrics focus on different aspects of quantum computers, which can favor one technology over another. Therefore, creating a universal metric is a challenging task. Examples of metrics that gained some traction are the Quantum Volume \cite{quantumvolume}, Circuit Layer Operations Per Second (CLOPS) \cite{CLOPS} and randomised benchmarking \cite{randobenchmarking}. Although such component-level benchmarks are important, they lack some comprehensibility, which becomes more important once we get closer to the wide applicability of quantum computers.

Recently, application-oriented benchmarks were introduced \cite{QED-C}, \cite{Qscore}. These metrics estimate the sizes of actual problems that can be solved with a quantum device. By focusing on applications, these metrics are naturally useful, hardware-agnostic and comprehensive. But, depending on the problem, they are not necessarily easily computable or scalable. By devising application-oriented benchmarks in a presentable and scalable way, such metrics promise to clearly measure the realistic performance of a quantum computer. The Q-score by Atos \cite{Qscore} and the metric by the QED consortium \cite{QED-C} are examples of application-oriented metrics. A more comprehensive discussion on recent metrics can be found in the table from the Boston Consultancy Group (BCG) white paper \cite{BCG}.

In this work, we will focus on the Q-score. The main advantage of the Q-score is that it predicts the efficacy of quantum devices in a real-world application. Furthermore, the Q-score can be computed for any computational device and hence allows a direct comparison between quantum and classical computers. The Q-score tests the entire computation stack of a computational setup, which is often not limited to the quantum device. On the one hand, this ensures that this metric measures the performance of a quantum device in practice, as quantum devices are always part of a larger computational workflow. However, this also makes it harder to compare the quantum aspect of various devices. Although the Q-score is proposed for a useful and representative optimisation problem, it does not say much about the performance on the variety of other interesting problems that are relevant for quantum computers.

Atos proposes to approximately solve the Max-Cut problem for large graphs as a representative application for quantum devices \cite{Qscore}. They define the Q-score of a computational approach as the size of the largest graph for which the setup can approximately solve the Max-Cut problem with a solution that significantly outperforms a random guessing algorithm. For general purpose quantum computers, they propose to use the Quantum Approximate Optimisation Algorithm (QAOA) \cite{FahriGoldstoneGutmann:2014}, as they view it as representative of practical needs of industry and challenging for current hardware. As the Max-Cut problem can be formulated as a Quadratic Unconstrained Binary Optimisation problem (QUBO), it can be solved on a quantum annealer as well. Due to the relatively small size of gate-based quantum computers, we expect a fairly small Q-score for these devices. Quantum annealers are already quite large, which raises the following questions:

\begin{enumerate}
\item What Q-score does a state-of-the-art quantum annealer achieve?
\item How do these Q-scores compare to classical algorithms for the Max-Cut problem?
\item What Q-score does a hybrid combination of a quantum annealer and a classical algorithm achieve? 
\end{enumerate}

In this paper, we give an answer to these three questions by reporting the Q-score of the various approaches by D-Wave.
Since Atos, at the time of writing, has not yet published any Q-scores of their own, the results reported in this paper give a first indication of what kind of Q-scores we can expect for quantum devices. This will then give an indication of how quantum devices perform on optimisation problems compared to classical devices. In addition, it will give an indication on when quantum computers could outperform classical computers on solving optimisation problems, potentially even achieving quantum advantage. Recall that quantum advantage is defined as demonstrating a problem solvable by a quantum computer for which no classical device can find a solution in a feasible amount of time.

We investigate the D-Wave Advantage system, D-Wave's latest and largest device as well as their previous generation device, the D-Wave 2000Q \cite{Tech_report_dwave}. In addition, we report the Q-score of their out-of-the-box hybrid approach \cite{Tech_report_dwave_hybrid} and of their implementations of the classical algorithms tabu search and simulated annealing. We start our work with a background on the Q-score and quantum annealing in Section~\ref{sec:background}, after which we present our results of the Q-score in Section~\ref{sec:results}. We finish our work with a discussion and directions for further research in Section~\ref{sec:discussion}.

\section{Background}
\label{sec:background}
\subsection{Atos Q-score}
In 2021, Atos introduced the Q-score \cite{Qscore}. They set out to define a metric that is simultaneously application-centric, hardware-agnostic and scalable. Note that with these three characteristics, the Q-score inherently has the four desired properties for a quantum metric mentioned above. Furthermore, the Q-score is given by a single number, where a higher Q-score implies that a device has a higher performance. This makes the Q-score accessible to the public as well. 

They propose approximately solving the Max-Cut problem for large graph sizes as a representative application of a quantum device. Given a graph $G$ of size $N$, i.e. with $N$ vertices, a cut $P$ is a partitioning of the nodes in two sets. The cost of a cut is the total number of edges between these two sets. The Max-Cut problem entails finding a cut that maximises this cost. They consider ($N$, $\frac12$)-Erd\"os-R\'enyi graphs, which are graphs of size $N$ where each edge is added with probability $\frac12$. By trying to find a maximal cut for multiple graphs of size $N$ on a certain device, we can compute the average best cut $C(N)$ for graphs of size $N$ obtained by this device.
We define the Q-score to be the highest $N$ for which a setup can find a solution that is on average significantly better than a random cut. Specifically, the largest $N$ for which 
\begin{equation}
\label{eq:beta}
    \beta(N) = \frac{C(N) - \frac{N^2}{8}}{C_{\rm max}(N) - \frac{N^2}{8}},
\end{equation}
is larger than $\beta^*=0.2$ is the Q-score\footnote{Atos acknowledges that the constant $\beta^*=0.2$ is chosen somewhat arbitrarily but they have picked it so that a noiseless quantum simulator running the QAOA algorithm with depth 1 achieves an infinite Q-score for this cut-off.}. Here $\frac{N^2}{8}$ is the expected cost of a random cut, and $C_{\rm max}(N)$ is the cost of the optimal cut for a graph of size $N$.
We can see that on average $\beta(N)=1$ for a setup that achieves the optimal cut and $\beta(N)=0$ for a setup that produces a random cut. It hence makes sense to find the maximum value of $N$ for which $\beta(N)$ is relatively high.
To make this computation scalable, Atos proposes the approximation $C_{\rm max}=\frac{N^2}{8} + 0.178 N^{\frac{3}{2}}$, where the 0.178 comes from a numerical fit of the optimal classical solution to Max-Cut in the range $N\in[5,40]$. Note that this might induce errors, especially at low graph sizes $N$.

Atos originally proposed to solve the Max-Cut problem for universal quantum devices with the Quantum Approximate Optimisation Algorithm (QAOA) \cite{FahriGoldstoneGutmann:2014}, as they view it as representative of practical needs of industry and challenging for current hardware. However, due to the hardware agnosticism and the scalability, the Q-score can be computed for any computing device and by with other methods.

To compute the Q-score, one generates $M$ Max-Cut problem instances of size $N$ at random and then computes the average cost $C(N)$ over these $M$ problems for the algorithm in question. The constant $M$ should be chosen large enough to mitigate the effect of stochasticity.

\subsubsection*{Q-score constraints}

The Q-score leaves some practical degrees of freedom, such as the computing time, optimisation methods, number of layers $p$ in QAOA, etc. Here, we explicitly state our constraints for computing the Q-score. Note that other choices for these constraints are possible and will likely lead to different Q-scores. Therefore, it is important to report the Q-score together with these constraints, most notably the time limit.

Firstly, we fix the maximum computation time at {\bf 60 seconds}, which is chosen so that both quantum and classical devices are expected to solve the Max-Cut problem for at least some problem sizes $N$. This time includes the entire computation starting from either a QUBO or a graph representation of the problem (see Section \ref{sec:annealers} for a definition of QUBOs). Note that it thus does not include the time to generate the problem instances. It does however include the embedding time, i.e. the time to translate the given QUBO to a hardware specific formulation. This is in the spirit of the Q-score as it is meant to benchmark the entire computation stack.

Secondly, we set the number of random graphs instances $M$ for each graph size to 100. In addition, the value of $\beta^* = 0.2$ was chosen by Atos. Therefore we do not just report the Q-score of the different devices, but also the $\beta$ vs $N$ graphs. 

When the algorithm does not produce a meaningful result (i.e. a cut), we assign it $\beta=0$, the value that is obtained by randomly guessing a solution. This can happen for a variety ovf reasons, such as a quantum device having insufficient resources to solve problems of size $N$, or a quantum device not finding a way to translate the graph into the required problem formulation within the time limit. When an approach does not admit a time limit, its Q-score will be the largest $N$ for which it can on average find an exact solution within 60 seconds.

We consider out-of-the-box solvers as a reproducible and first indication of the Q-score. In practice, solvers can benefit from a problem-specific optimisation to obtain higher Q-scores, e.g. optimal time distribution among algorithm steps. We chose to avoid such optimisations to allow the most neutral and fair comparison of the different approaches. This means that we restrict ourselves to supplying $M=100$ random graphs of size $N$, limiting the solving time to 60 seconds per graph and running the solvers with default parameters.

\subsection{Quantum Annealing and D-Wave}\label{sec:annealers}
There are multiple flavours of quantum computer of which two are currently most researched, namely gate-based devices and quantum annealers. In this work we will only consider quantum annealers. Quantum annealers are devices that can solve optimisation problems that are submitted in the form of a Quadratic Unconstrained Binary Optimisation problem (QUBO). A QUBO is an expression that is a sum of terms where each term is either constant, linear or quadratic in the binary variables $x_i$. When solving a QUBO, the aim is to find binary variable assignments that minimise the QUBO. Many optimisation problems can be formulated as a QUBO, so that the minimal value of the QUBO translates to the optimal solution of the optimisation problem \cite{QUBO_problems}. The Max-Cut formulation is shown to be equivalent to the QUBO formulation in section 3.2 of \cite{QUBO_problems}. Together, this implies that the Max-Cut problem can be seen as a representative problem instance for the class of optimisation problems.

Quantum annealers approximately solve QUBOs by using principles from quantum mechanics. A quantum annealer is a quantum device with some number of qubits, which are connected via a certain topology. Given a QUBO, the quantum annealer identifies each variable with a number of qubits, so that terms in the QUBO correspond to connections on the quantum annealer. This identification of QUBO with qubits is called an embedding. Quantum annealers prepare qubits in an initial uniform superposition and then let them evolve under a problem-specific Hamiltonian. If this evolution is slow enough, the qubits are likely to end up in the ground-state or other low energy-states of this Hamiltonian. When encoded correctly, the ground-state corresponds to the minimum of the QUBO and the other low energy-states correspond to values close to this minimum. By measuring the qubits after the evolution, we find values that approximately minimise the QUBO.

\subsubsection*{D-Wave Devices and Solvers}
Currently, D-Wave offers three types of solvers: quantum annealers, classical solvers, and hybrid solvers \cite{D-wave}, \cite{qbsolv}, \cite{neal}. These solvers are all constructed to solve QUBOs. At the time of writing, their biggest quantum devices are the Advantage device, which has 5640 qubits, where each non-boundary qubit is connected to 15 other qubits and the 2000Q device, which has 2048 qubits, where each non-boundary qubit is connected to 6 other qubits. Both devices have a certain topology on their qubits, that describes how the qubits are connected. The topologies of the D-Wave devices are designed to maximise the amount of different QUBOs that can be solved on these devices.

In addition to these quantum devices, we wish to compute the Q-score of some classical algorithms to compare them to the performance of the D-Wave. There is a wide variety of classical algorithms that solve the Max-Cut problem (see for example the overview in \cite{classical_max_cut}), however most of these algorithms do not mention the time performance and even fewer have public source code for their implementation. As an exception, the tabu search implementation in \cite{tabu_search_paper} does mention its time performance. However, this implementation requires time in the magnitude of hours to solve problems of size larger than 5,000. There are some companies that offer classical solvers out of the box that can be used for the Max-Cut problem, such as D-Wave, LocalSolver, CPLEX and Gurobi. Two examples of classical solvers from D-Wave are Qbsolv and simulated annealing. Qbsolv starts by splitting the QUBO into smaller pieces and then solves the indivdual pieces with a tabu search algorithm. Simulated annealing is a local search algorithm that looks for minima by having a slow decrease in the probability of accepting worse solutions as the solution space is explored. Both of these methods appear in lists of best classical approaches to solve the Max-Cut problem \cite{classical_max_cut}. Hence, we have chosen Qbsolv and simulated annealing as classical algorithms to compare the D-Wave quantum devices to. 

Last but not least, D-Wave offers a hybrid solver, which combines the strengths of both classical and quantum solvers to solve QUBOs \cite{Tech_report_dwave_hybrid}. The hybrid solver uses D-Wave's Advantage device for the quantum part, while it uses a collection CPU and/or GPU platforms provided by Amazon Web Services (AWS) for the classical part. Because of this combination of strengths, the hybrid solver promises to have the best performance at solving QUBOs. As D-Wave's technical report shows, the hybrid solver outperforms many of the best classical algorithms at solving QUBOs, especially at larger problem sizes \cite{Tech_report_dwave_hybrid}. However, D-Wave does not provide information about the distribution of quantum and classical computing time and power. Because of this, this solver is more of a black box than the other solvers.

We have implemented this setup using python and used this code to obtain the results below. For transparency, we have made our code at https://github.com/TNO-Quantum/qscore$\_$dwave. It consists of several scripts, allowing to solve random Max-Cut instances on any of the five solvers and compute the corresponding beta.

\section{Results}
\label{sec:results}
To compute the Q-score, we generate 100 random ($N,\frac{1}{2}$)-Erd\"os-R\'enyi graphs for increasing problem sizes $N$ with the NetworkX python library. Next, we convert these graphs into QUBOs and (approximately) solve the Max-Cut problem for these QUBOs by finding binary variable assignments with low values for the QUBO. We do this by supplying these QUBOs to D-Wave's solvers provided by the D-Wave's open source library \cite{ocean}. The time constraint of 60 seconds is implemented in different ways depending on the solver, each of which is explained separately below. To apply the solvers in an out-of-the-box fashion, we do not supply or change any other input variables. Having obtained results on all 100 graphs, we compute the average best cut found $C(N)$ and compute $\beta(N)$ accordingly with the formula above. We increase $N$ until we find a value of $N$ for which $\beta(N)\leq 0.2$.

For the classical algorithms we use a modest server with an Intel Core i7-7600U CPU with a clock speed of 2.80GHz and 12 GBs of RAM.

\subsection{Q-scores of D-Wave's classical solvers}
\label{sec:classical_results}
\begin{figure}
\centering
\begin{subfigure}{\columnwidth }
  \centering
    \includegraphics[width=\columnwidth]{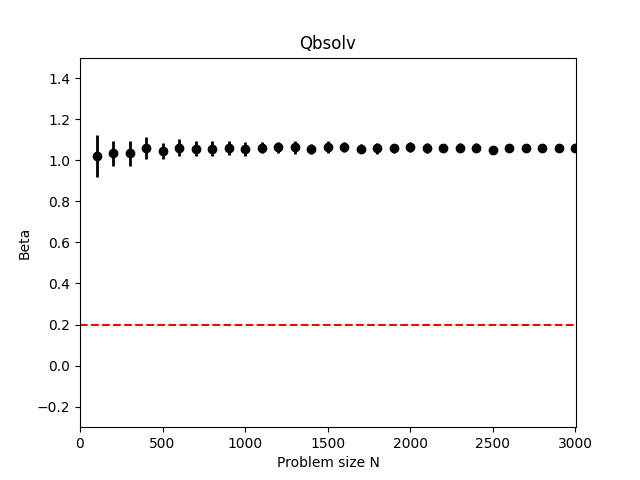}
    \caption{Q-score's $\beta$ vs problem size $N$ for tabu search by Qbsolv.}
    \label{fig:Qbsolv_beta}
\end{subfigure}
\begin{subfigure}{\columnwidth }
  \centering
    \includegraphics[width=\columnwidth ]{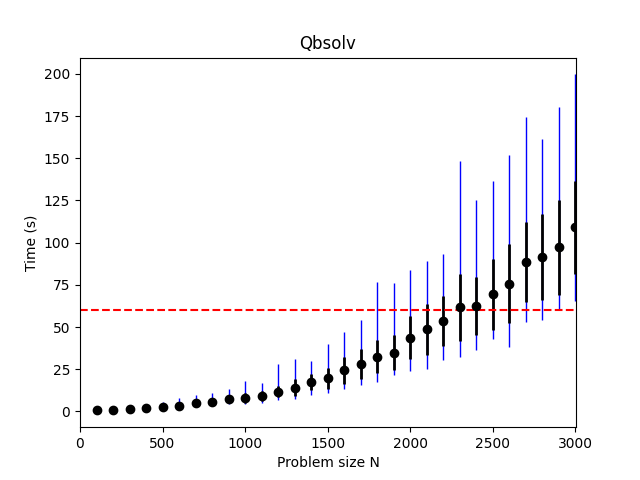}
    \caption{Time vs problem size $N$ for tabu search by Qbsolv.}
    \label{fig:Qbsolv_time}
\end{subfigure}%
\caption{Tabu search by Qbsolv results.}
\label{fig:tabu}
\end{figure}

\begin{figure}
\centering
\begin{subfigure}{\columnwidth}
  \centering
    \includegraphics[width=\columnwidth]{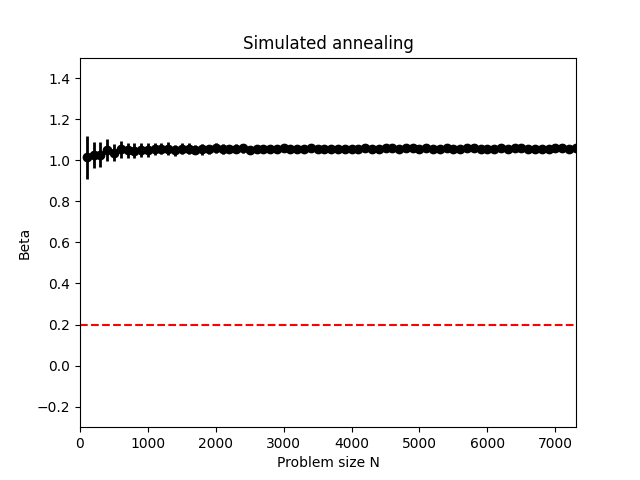}
    \caption{Q-score's $\beta$ vs problem size $N$ for simulated annealing.}
    \label{fig:SA_beta}
\end{subfigure}
\begin{subfigure}{\columnwidth}
  \centering
    \includegraphics[width=\columnwidth]{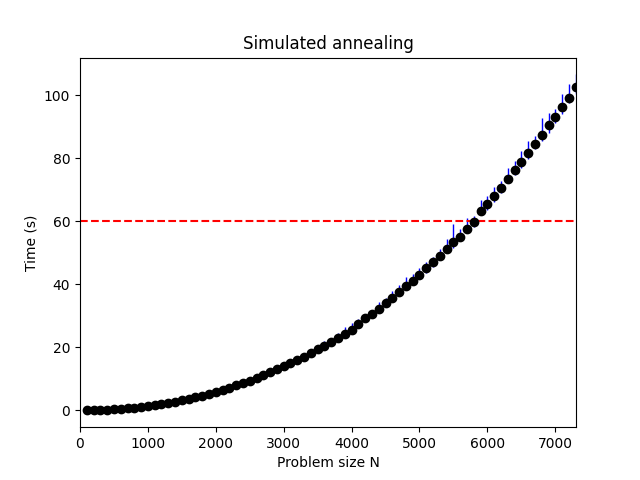}
    \caption{Time vs problem size $N$ for simulated annealing.}
    \label{fig:SA_time}
\end{subfigure}%
\caption{Simulated annealing results.}
\label{fig:SA}
\end{figure}

Here we report the Q-score of two out-of-the-box classical solvers provided by D-Wave's open source library. We use tabu search as provided by the Qbsolv class \cite{qbsolv} as well as simulated annealing as provided by D-Wave's neal package \cite{neal}. Fig. \ref{fig:tabu} and Fig. \ref{fig:SA} show $\beta$ and the running time plotted against the problem sizes N.
The blue vertical lines indicate the minimum and maximum time recorded. The black lines represent the statistical standard deviation\footnote{Since random graph instances can have a different optimal cost, we expect systematic and statistical error. This effect will inflate the error bars for low $N$. However, it does not have a big effect on the estimation of the Q-score so we simply treat all error as statistical standard deviations.}.
For both approaches we start at $N=100$ and increase $N$ by 100 nodes each time. We terminate the computation when the average time for a graph size exceeds 100 seconds on our server. We see that these approaches yield good cuts with $\beta \approx 1$ in the entire range. 

To impose the time constraint of 60 seconds, one should note that there is a degree of freedom is how much time is spent in inner and outer loops. This allows us to cut off the algorithm at 60 seconds in different ways. Although Qbsolv has a {\it timeout} parameter, it works by comparing the expired time with the {\it timeout} at the end of every outer loop. This results in problem instances where the total time taken exceeds 60 seconds, even for {\it timeout} parameters less than a second. Similarly, to terminate simulated annealing at 60 seconds inherently requires an optimisation. Rather than finding the ideal way to set this cut-off, we report the average time taken per graph instance and report the Q-scores to be the respective largest graph sizes for which the algorithms take less than 60 seconds on average. This allows us to read off the Q-score without having to do specific optimisations.

We see that simulated annealing outperforms Qbsolv, as it achieves a Q-score of 5,800 versus the 2,300 achieved by Qbsolv search within the 60 second time limit.

\begin{figure}
    \centering
    \includegraphics[width=\columnwidth]{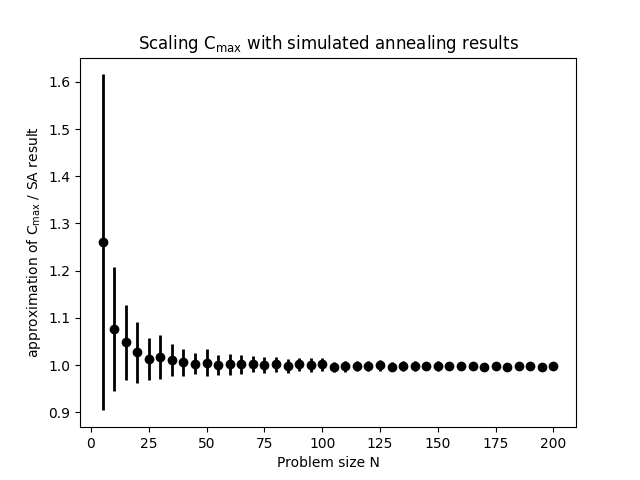}
    \caption{Ratio of the approximation for $C_{\rm max} \approx \frac{N^2}{8} + 0.178 N^{3/2}$ and the mean cost of simulated annealing.}
    \label{fig:correction}
\end{figure}

The results yield a $\beta$ slightly higher than 1. This is due to the fact that the approximation of $C_{\rm max}$ in  Eq. (\ref{eq:beta}) is not exact. Comparing the approximation $C_{\rm max} \approx \frac{N^2}{8} + 0.178 N^{3/2}$ to the average best cut found by simulated annealing, we see that especially for low $N$ the approximation is not perfect. See Fig.~\ref{fig:correction} for the ratio between the simulated annealing result and the approximation as a function of $N$ at low $N$. In later results we will compute $\beta$ in two ways, namely using for $C_{\rm max}$ both the approximation as well as the average best cut found by simulated annealing.

\subsection{Q-scores of D-Wave hardware}

Here, we report the Q-score of the D-Wave Advantage device, as well as the previous generation device, the 2000Q. The D-Wave interface allows us to set a maximum time of 60 seconds per problem instance. For both devices we start at $N=5$ and increment $N$ in steps of five. We stop this process when $\beta$ is clearly below 0.2 or no solution is found within the allowed time.

\begin{figure}
\centering
\begin{subfigure}{\columnwidth}
  \centering
  \includegraphics[width=\columnwidth]{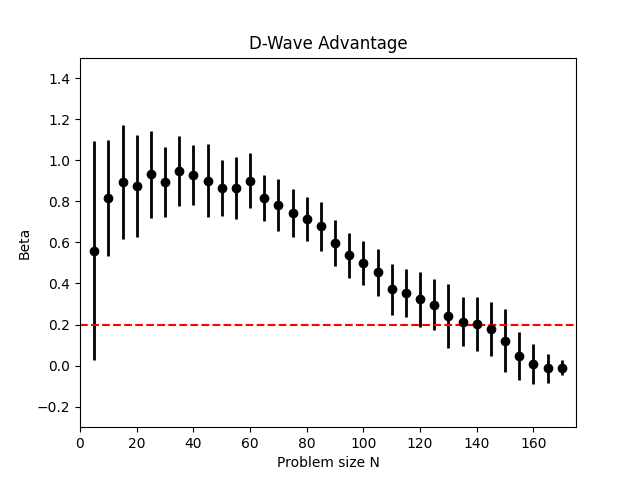}
  \caption{Q-score's $\beta$ vs problem size $N$ for the D-Wave Advantage QPU.}
  \label{fig:DWave_QPU}
\end{subfigure}
\begin{subfigure}{\columnwidth}
  \centering
  \includegraphics[width=\columnwidth]{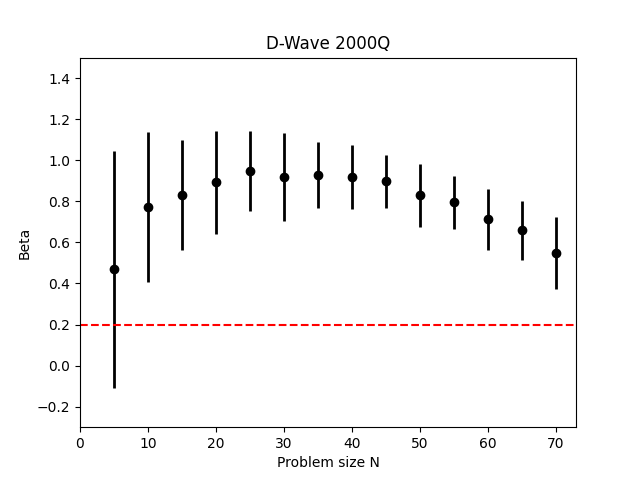}
  \caption{Q-score's $\beta$ vs problem size $N$ for the D-Wave 2000Q QPU.}
  \label{fig:DWave_2000Q}
\end{subfigure}
\caption{Q-score's $\beta$ vs problem size $N$ for the D-Wave quantum annealers with an approximated $C_{\rm max}$ in Eq. (\ref{eq:beta}).}
\label{fig:qpu_nanis0}
\end{figure}

In Fig. \ref{fig:qpu_nanis0} the $\beta$ vs problem size $N$ graph is plotted for the D-Wave Advantage and the D-Wave 2000Q in Fig.. For the D-Wave Advantage, $\beta=0.2$ is crossed at $N=140$, which means that we find a Q-score of $140$. 

The $\beta$ vs $N$ graph of the D-Wave 2000Q device only produces an approximate solution up to $N=70$. For larger graphs, the system is unable to find an embedding of the resulting QUBO onto the hardware. This inability to find an embedding is time independent, so our 60 second limit is not a factor. Following our definition we set $\beta=0$ for $N>70$ and hence report a Q-score of 70.

\begin{figure}
\centering
\begin{subfigure}{\columnwidth}
  \centering
  \includegraphics[width=\columnwidth]{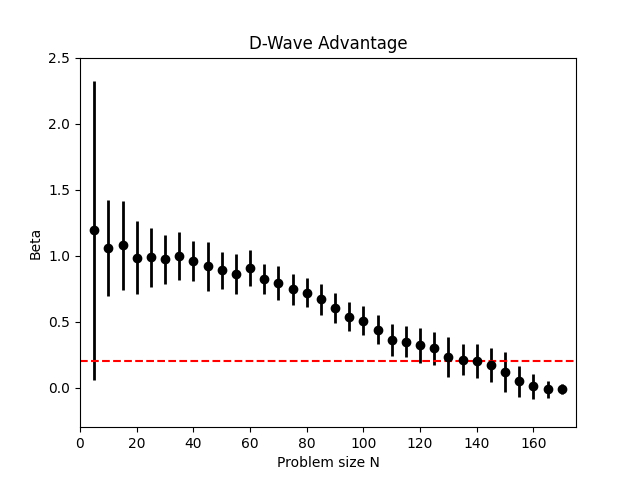}
  \caption{Q-score's $\beta$ vs problem size $N$ for the D-Wave Advantage QPU with the simulated annealing result as $C_{\rm max}$.}
  \label{fig:adv_scaled}
\end{subfigure}
\begin{subfigure}{\columnwidth}
  \centering
  \includegraphics[width=\columnwidth]{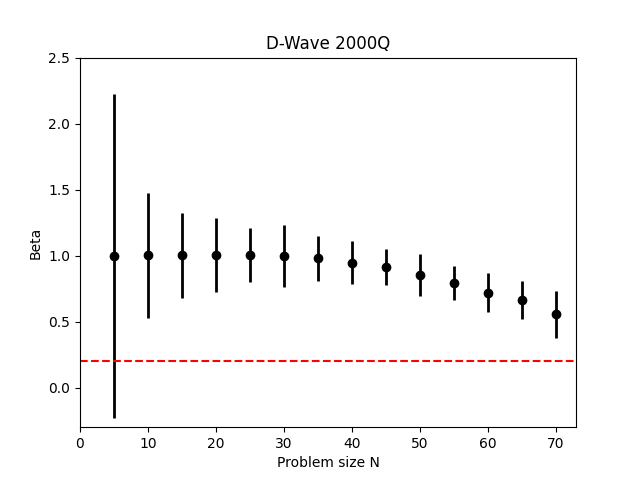}
  \caption{Q-score's $\beta$ vs problem size $N$ for the D-Wave 2000Q QPU with the simulated annealing result as $C_{\rm max}$.}
  \label{fig:2000Q_scaled}
\end{subfigure}%
\caption{Q-score's $\beta$ vs problem size $N$ for the D-Wave quantum annealers with the simulated annealing result as $C_{\rm max}$ in Eq. (\ref{eq:beta}).}
\label{fig:scaled_hardware}
\end{figure}

For both devices, we find $\beta < 1$ for low $N$. This is due to the approximation of the optimal cut, $\frac{N^2}{8} + 0.178 N^{3/2}$, not being very accurate for low $N$ as seen in Fig.~\ref{fig:correction}.
Replacing the approximation of $C_{\rm max}$ by the simulated annealing result, we find the expected result of $\beta \approx 1$ at low $N$, see Fig.~\ref{fig:scaled_hardware}.

\subsection{Hybrid solver}
\begin{figure}
    \centering
    \includegraphics[width=\columnwidth]{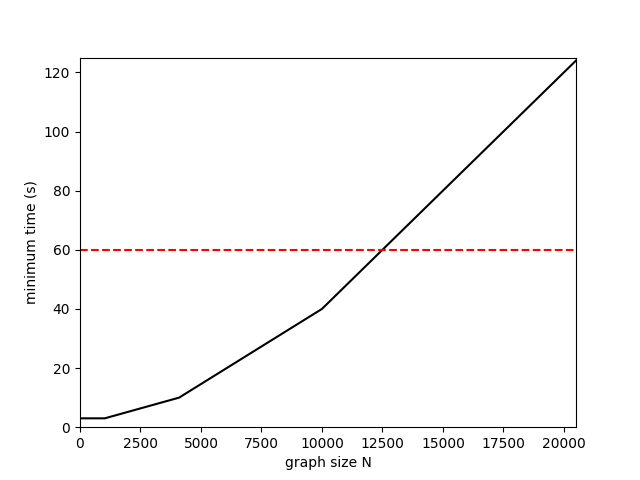}
    \caption{Minimum time required to pass to the hybrid binary quadratic model solver.}
    \label{fig:hybrid_time}
\end{figure}

D-Wave's hybrid binary quadratic model solver is used out of the box for Max-Cut problems. Just as on the D-Wave hardware, the D-Wave interface allows us to set a maximum time of 60 seconds per problem instance. We ran the hybrid solver from $N=100$ until $N=5{,}000$ in steps of 100 and found $\beta \approx 1$ in the entire range. In addition we ran the algorithm for $N=12{,}500$ and still found $\beta \approx 1$. At this point we were confronted with a hard-coded time limitation of the hybrid solver \cite{time_limit}. It requires a minimum time per problem instance given by a linear interpolation between some fixed points, see Fig.~\ref{fig:hybrid_time}. Based on this plot we expect the effective Q-score to be about 12,500. However, due to this hard-coded time limitation we cannot see the result for larger graphs with our chosen time limit. It might very well be that we run into the same problem as with our classical algorithms, namely that it will not complete a loop within the given time for larger graphs.

\section{Discussion}
\label{sec:discussion}
We show an overview of the Q-scores of the various approaches in Table \ref{tab:overview}.

\begin{table}
    \centering
    \caption{Q-scores of D-Wave's QUBO solvers with a 60 seconds time limit.}
    \begin{tabular}{l|r}
        Approach &  Q-score \\
        \hline
        Tabu search & 2,300\\
        Simulated annealing & 5,800\\
        D-Wave Advantage & 140\\
        D-Wave 2000Q & 70\\
        Hybrid solver & 12,500
    \end{tabular}
    \label{tab:overview}
\end{table}

If we compare the classical, fully quantum, and hybrid approaches, we see that the classical solvers significantly outperform the approaches with only quantum computations. In turn, the hybrid approach outperforms both given the fixed maximum time. From a business perspective it makes sense that the hybrid solver by D-Wave outperforms their free classical approaches. It would therefore be interesting to compare these Q-scores to classical approaches by other companies, such as LocalSolver, CPLEX and Gurobi. We tried to compute the Q-score using other classical algorithms, however found no improvement. 

We should mention that this comparison between hybrid and classical solvers is somewhat skewed. We ran the classical solvers on a modest personal server. The hybrid solver however uses a collection of CPU and/or GPU platforms provided by Amazon Web Services for its classical computations. We expect these CPU and GPU platforms to outperform our modest server. For a better comparison, one should run both the hybrid solver and the classical solvers on the same classical hardware backend. 

Another interesting direction for future research is lifting the hard-coded time limits for the hybrid solver, which seems to lead to higher Q-scores for this solver. This requires opening the black-box implementation, which we cannot do. 

With these results we can to conclude that hybrid devices already outperform classical ones at solving a representative optimisation problem. The current difference is however not yet significant enough to speak of real quantum advantage on optimisation problems. Considering the focus of the Q-score on optimisation problems, it should be noted that this metric need not give any indication of the usefulness of a quantum device on other applications. Nonetheless, it will be exciting to see whether quantum developers can potentially extend this small advantage into a full quantum advantage in the future, especially considering that quantum computing is still in its infant years.

Reporting the Q-score together with a time limit worked well for D-Wave's Advantage device up to $N=180$ and the 2000Q device up to $N=70$. However, for the classical and hybrid approaches we found no relation between $\beta$ and $N$ in which $\beta$ degraded over time. We expected this for exact solvers and it turns out to also be the case for out-of-the-box use of approximate algorithms. Although we can still report a Q-score based on the time spent, we believe the classical and hybrid results are less in line with the spirit of the Q-score since they do not find approximate solutions but only exact optimal solutions. To address this issue, one could instead consider classical solvers that do allow approximate solutions when time is limited. This would allow a comparison of quantum and classical devices which is more in line with the spirit of the Q-score.

It is clear that the approximation used to compute $\beta$ is not exact. We consistently find $\beta\approx 1.06$ instead of $\beta=1$ for large $N$. For low $N$ the approximation is not needed, since optimal classical solutions can be found easily. We believe it is better not to use the approximation for $N<100$, as at low $N$ this introduces large uncertainties in the $\beta$ parameter. If one uses a standardised set of graphs, rather than randomly generated graphs this error could be significantly reduced since the exact optimal cut would be known. However, since even with the approximation the $\beta$'s are still sufficiently higher than the boundary of $\beta*=0.2$, this does not affect the eventual Q-scores.

We reported $\beta=0$ in cases where no embedding on the device can be found. By setting $\beta=0$ we effectively benchmark the device with an additional step of generating a random cut when no embedding is found. One could argue this is not as out-of-the-box as other algorithms. However, since the Q-score aims to benchmark the entire setup, not just the quantum hardware, this step would be a trivial addition. A much harsher alternative would be to set the cost to zero when no embedding is found, though as these occurences are rare, the effect on the Q-score is limited.

As mentioned in the introduction, the Q-score can also be computed for gate-based devices. However, at the moment of writing, these devices are much smaller in size than the quantum annealers provided by D-Wave. That is why we expect a state-of-the-art gate-based device to have a signficantly lower Q-score than the 140 we found for the quantum annealers. It is interesting however to compute the Q-score for a gate-based device as well so that we can explicitly compare these two paradigms of quantum computing. Looking at the currently available topologies, gate-based quantum computers will not yet beat the Q-scores of the quantum annealers, let alone the Q-scores of classical and hybrid algorithms. A moment to look forward to is when gate-based devices reach similar performances as quantum annealers or classical algorithms for optimisation tasks. 

In conclusion, classical devices currently outperform state-of-the-art quantum annealers on optimisation problems, while a hybrid approach combining the two seems to outperform classical approaches. These differences are however not yet significant enough to give strong hints towards quantum advantage.

\bibliographystyle{IEEEtran}
\bibliography{main.bib}

\end{document}